\documentclass[aps,pre,twocolumn,showpacs]{revtex4-1}
\usepackage{graphicx}
\usepackage[colorlinks=true,linkcolor=blue,%
linkcolor=blue,pagecolor=blue,citecolor=blue,%
urlcolor=blue,anchorcolor=black,bookmarksnumbered=true,bookmarksopen=true]{hyperref}%
\renewcommand{\frac}[2]{\displaystyle{#1 \over #2}}
\begin{document}
\title{ Nonviscous motion of a slow
particle in the dust crystal under
microgravity conditions}
\author{D.~I.~Zhukhovitskii} \email{dmr@ihed.ras.ru}
\homepage{http://oivtran.ru/dmr/}
\author{V.~E.~Fortov}
\author{V.~I.~Molotkov}
\email{molotkov@ihed.ras.ru}
\author{A.~M.~Lipaev}
\author{V.~N.~Naumkin}
\affiliation{Joint Institute of High Temperatures, Russian
Academy of Sciences, Izhorskaya 13, Bd.~2, 125412
Moscow, Russia}
\author{H.~M.~Thomas}
\author{A.~V.~Ivlev}
\email{ivlev@mpe.mpg.de}
\author{M.~Schwabe}
\author{G.~E.~Morfill}
\affiliation{Max-Planck-Institut f\"{u}r extraterrestrische
Physik, Giessenbachstrasse, 85748 Garching, Germany}
\begin{abstract}
Subsonic motion of a large particle moving through the bulk of a dust crystal formed by negatively charged small particles
is investigated using the PK-3 Plus laboratory onboard the International Space Station. Tracing the particle trajectories
show that the large particle moves almost freely through the bulk of plasma crystal, while dust particles move along
characteristic $\alpha$-shaped pathways near the large particle. In the hydrodynamic approximation, we develop a theory of
nonviscous dust particles motion about a large particle and calculate particle trajectories. A good agreement with
experiment validates our approach.
\end{abstract}
\maketitle
\section{\label{1} INTRODUCTION}

Complex or dusty plasmas \cite{Morfill09,2} are low temperature plasmas consisting of electrons, ions, neutral gas atoms,
and charged microparticles or dust particles. Due to the higher electron mobility, these particles acquire large negative
charges up to $10^4 e$, where $e$ is the electron charge. Thus, the particle ensemble is a strongly coupled classical
Coulomb system, in which plasma crystal, i.e., ordered structures of dust particles can be observed
\cite{1,3,4,5,6,Fortov05}. In experimental setups, dusty plasmas are usually studied in gas discharges at low pressures,
e.g., in radio frequency (RF) discharges \cite{1} or in striations of dc discharges \cite{7,8,EThomas}. In ground-based experiments,
the electric field present in a discharge compensates the force of gravity and plays the role of a trap for dust particle
structures. Three-dimensional structures obtained in RF and dc discharges are inhomogeneous ones. A large homogeneous bulk
of dusty plasma, which almost fills the entire discharge volume, can be observed under microgravity conditions either in
parabolic flights \cite{9,10,11,12,13} or onboard the International Space Station (ISS) \cite{9,14,15,16,17,18}.

Recently, self-excited dust--density waves caused by streaming ions were studied \cite{12}, and freezing and melting of dusty
plasma upon variation of the neutral gas pressure was investigated \cite{16}. The motion of a large particle inside the dust
crystal of smaller particles was reported in \cite{10,13,18}. It was demonstrated that supersonic motion is accompanied by
the formation of the Mach cones, causing noticeable perturbation of plasma crystals. There were also relatively slow
particles observed, moving with almost a constant velocity or even accelerating, which is incompatible with common notion of
the interaction between the particle and dust crystal. Such particles appear to move freely, without long-range distortion
of the ambient structure.

Using the PK-3 Plus laboratory onboard the ISS we have investigated motion of the dust particles in the vicinity of moving
large particle. Such relatively slow particles are accelerated spontaneously on the outside of the particle cloud in RF
discharge, penetrate into the dust cloud, and move inside it. We have made a series of snapshots of this motion in the plane
of a laser sheet with a high-resolution camera, which allowed us to trace the trajectories of individual particles. Motion
of small particles caused by the passage of a large particle has already been studied and used as a source of valuable
information in the course of a ground-based experiment \cite{21}, where heavy particles were dropped down to the cloud of
dust particles levitating in the plasma sheath region of a RF capacitive planar discharge. Trajectories of particles
suspended in low pressure glow discharges have also been studied in \cite{22}. Dust particles were arranged in chains, and a
heavy particle falling between neighboring dust chain bundles caused the elliptical motions of the background dust
particles.

Superposition of successive frames recorded in our experiment reveals that many more dust particles in a close neighborhood
of a moving particle circumscribe typical $\alpha$-shaped trajectories, while other particles remain almost at rest. We
interpret this motion as a nonviscous flow about a large particle,
for which we used a hydrodynamic approach. This approach is widely used as applied to the collective motion of dust
particles \cite{2,23,24,Ivlev07}. Based on classical solution for the velocity field \cite{25}, we integrate it to obtain
streamline pathways for individual dust particles. A good agreement between recorded and calculated trajectories validates
our basic assumption. Since a nonviscous motion is associated with zero momentum transfer, the force acting on a large
particle vanishes in the idealized hydrodynamic model. If the particle interact only with the dust particles in its close
neighborhood, as in the case of a nonviscous flow, the estimate of the direct particle--dust interaction force suggests that
it is not greater than other major forces operating in low-pressure gas discharges (such as the ion
drag and electric forces).

The paper is organized as follows. In Sec. II, we describe
the experimental setup and results of tracing the dust
particle motion caused by the passage of a large particle. A
hydrodynamic approach is developed in Sec. III, where
both numerical and analytical solutions for dust particle
trajectories are obtained. Results are summarized in
Sec. IV.

\section{\label{s2} EXPERIMENTAL SETUP AND
RESULTS}

Here, we present an observation of a subsonic nonviscous motion of a large particle in the bulk of a three-dimensional
complex plasma in the PK-3 Plus laboratory onboard the ISS. Details on the setup can be found in \cite{17}. The heart of
this laboratory consists of a capacitively coupled plasma chamber with circular electrodes of 6 cm diameter and 3 cm apart
(see Fig.~\ref{f1}). Microparticles can be injected into the main plasma with dispensers. The particles then form a cloud
around the center of the chamber, typically with a central void caused by the ions streaming outwards. Generally, some
larger particles are present in the chamber as well. The origin of these particles is not yet understood; these might be,
for instance, agglomerates or larger particles left over from previous experiments (not removed during the cleaning
procedure). They normally accumulate themselves at the periphery of the particle cloud, because of the dependence of the
ratio of the electric and ion drag force on the particle diameter \cite{27}. Sometimes, one of these larger particles gets
sporadically accelerated and penetrates into the cloud -- we shall term it projectile. The reason for such behavior remains
a puzzle (for instance, it might be caused by a laser-induced rocket force \cite{28}). Nevertheless, these particles can be
used as natural probes of the microparticle cloud \cite{29,30}, in the same way as particles injected on purpose
\cite{10,13,31}. Projectiles enable various active experiments to be performed in a wide range of testing parameters, thus
providing the opportunity to reveal new features of strongly coupled complex plasmas. In particular, velocities of
projectiles can vary significantly, and when they move through a cloud of background microparticles at supersonic velocities
they excite Mach cones \cite{18}. On the other hand, obvious disadvantages of experiments with projectiles is that their
velocities cannot be controlled.

In what follows, we discuss in detail one example of a slow (subsonic) motion of a projectile. The experiment was performed
during the 13th mission of PK-3 Plus on the ISS. Argon was used as buffer gas at a pressure of 10 Pa, and the main
microparticle cloud was composed of melamine-formaldehyde particles with $2.55~\mu$m diameter. The projectile was
accelerated from the side and penetrated into the main cloud, moving almost horizontally from the left to the right (see
Fig.~\ref{f2}a, phase I). To determine the size of the projectile, this experiment was compared 
with others performed with larger particles, which allowed us to 
conclude that the projectile radius was $a_p=7.5~\mu$m. During its path towards the void, the velocity of the projectile varied, decreasing from 80 mm/s to 37 mm/s.
Inside the void, the horizontal motion further slowed down. The particle was then accelerated upwards (see Fig.~\ref{f2}a,
phase II), where it again penetrated into the microparticle cloud (see the discussion of microparticle trajectories in
Ref.~\cite{18}). In the region above the void, the particle was slower than before. It did, however, push the microparticles
away to clear its path. Figure~\ref{f2}b shows the projectile motion during phase II. The microparticles that were pushed
away moved in vortices around the probe particle. The projectile accelerated while traveling upwards through the
microparticle cloud, so that its velocity increased from 7 to 14 mm/s. This value is still lower than the speed of sound.

In the experiments, we monitored the motion of dust particles in the vicinity of the projectile, by manually determining its
position recorded by a high-resolution camera at 50 frames/s (a video frame is illustrated in Fig.~\ref{f2}b). The
horizontal and vertical resolution of the camera was $11.3~\mu$m per pixel and $10.3~\mu$m per pixel, respectively.

Experimental results on the motion of dust particles in the neighborhood of a slowly moving projectile are summarized in
Fig.~\ref{f3}. The projectile motion from the void to chamber wall is represented by frames from bottom to top in
Fig.~\ref{f3}a. Dust particles form a dust crystal, which can be represented as a set of the Wigner--Seitz cells around each
particle. We estimate the dust particle number density as $n_d = 3 \times 10^5 \;{\mbox{cm}}^{ - 3} $, hence the cell radius
is $\bar r_d = (3/4\pi n_d )^{1/3} = 9.3 \times 10^{ - 3} $ cm. We can define the coupling parameter of interaction between
the projectile and dust particles (or the scattering parameter) $\beta_{dp} = 2Z_p Z_d e^2 /\lambda M_d u^2 $ as the ratio
of characteristic Coulomb energy $Z_p Z_d e^2 /\lambda$ at the plasma (ion) screening length $\lambda$ (where $Z_p$ and
$Z_d$ are the charges of the projectile and dust particle, respectively, in units of the electron charge $e$) to the kinetic
energy of a dust particle $M_d u^2 /2$ in the reference frame co-moving with the projectile (where $M_d = 1.31 \times
10^{-11}$~g is the dust particle mass and $u\simeq 1$~cm/s is the projectile velocity). Based on the analysis of recent
melting experiments performed under identical conditions \cite{16}, we estimate $\lambda\simeq6\times10^{-3}$~cm,
$Z_d\simeq-1200$ and, assuming a linear scaling of charge with size, $Z_p \simeq-7000$. This yields $\beta_{dp}\simeq100$
and, hence, the estimate for the radius of strong interaction between the projectile and dust particles gives
\cite{Fortov05} $\simeq\lambda\ln\beta_{dp}\simeq2.7\times10^{-2}$~cm. This value turns out to be fairly close to the
average distance between the projectile and nearest dust particle neighbors, $\bar R\simeq3.3\times 10^{-2}$~cm (see
Fig.~\ref{f3}a).
Figure~\ref{f3}b shows successive phases of motion of an individual dust particle. Trajectories of such particles are
revealed by frame superposition (Fig.~\ref{f3}c). It is evident that many more particles that interact with the projectile
move along loops of similar form. The absence of loops in some regions can be accounted for by the following facts: first,
the projectile trajectory is not exactly parallel to the laser illumination plane as well as dust crystal planes, so one can
see only projections of dust particle trajectories. Second, visible pathways of the dust particles terminate at the points
where particles go out of the laser sheet. Third, as the distance from the projectile increases, dust particles come at rest
and become a part of the crystal. Upon such local crystallization, the particle can be pushed away from its initial
trajectory.

\section{\label{s3} THEORY OF NONVISCOUS DUST
PARTICLE MOTION}

Let us first estimate the major forces acting on the projectile, in order to enlighten the physics of interaction between
the projectile and the surrounding complex plasmas. We shall confine ourselves to treatment of the neutral drag force $F_n$,
the ion drag force $F_i$, the electric force $F_e$, and the force from dust particles $F_d$.

We start the analysis with the neutral drag force, for which we can provide the most accurate estimate. This force is
naturally pointed in the direction opposite to the projectile velocity. Assuming, as usual, diffuse scattering of neutrals
from the particle surface (with full accommodation) we obtain \cite{Fortov05}
\begin{displaymath}
F_n = \delta\frac{8\sqrt{2\pi}}3m_n n_n v_{Tn} a_p^2u \simeq 3\times 10^{-8}~{\rm dyn},
\end{displaymath}
where $\delta\simeq1.4$ is the accommodation coefficient, $n_n = 2.42 \times 10^{15}$~cm$^{-3}$ is the number density of
neutral gas at the pressure of 10~Pa, $m_n = 6.63 \times 10^{-23}$~g is the mass of argon atom, and $v_{Tn} = (T_n
/m_n)^{1/2} = 2.50\times 10^4 $~cm/s is the thermal velocity of neutrals.

To estimate the plasma-induced forces, we need to know the magnitude of a dc electric field, $E$. The field is generated due
to ambipolar diffusion and can be deduced from the scaling \cite{RaizerBook} $eEL\sim T_e$, where $L=3$~cm is the distance
between the RF electrodes (see Fig.~\ref{f1}) and $T_e = 3.5$~eV is the electron temperature (here and in what follows, the
temperature is expressed in the energy units). Then the electric force is
\begin{displaymath}
F_e = \left|{Z_p}\right|eE = \left|{Z_p}\right| T_e/L\sim 10^{-8}~{\rm dyn}.
\end{displaymath}
We note, however, that in dense clouds of dust particles, where the Havnes parameter is large, $E$ can be dramatically
reduced (by an order of magnitude or even more, see Refs~\cite{Land06,Sutterlin09}). Therefore, the obtained value only be
treated as a crude {\it upper-bound} estimate.

The ion drag force is determined by the ion drift velocity $u_i$. Assuming the mobility-limited ion drift we get $u_i =
eE/m_i \nu _i\sim T_e /(m_iv_{Ti}n_n\sigma_{in}L) \sim3\times 10^4$~cm/s, where $\nu _i = v_{Ti} n_n \sigma _{in} $ is the
ion--neutral collision frequency, $v_{Ti}(=v_{Tn})$ is the ion thermal velocity, and $\sigma _{in} \simeq 10^{-14}$~cm$^{-2}$
is the ion--neutral collision cross section. The magnitude of the ion drag force is determined by the ion scattering
parameter $\beta_{pi}=|Z_p|e^2/\lambda T_i$, which is of the order of 10 for our conditions. In this case, the force on the
projectile is dominated by the direct ion collisions (absorption) and can be estimated from the following formula
\cite{Fortov05}:
\begin{displaymath}
F_i\simeq\pi\lambda^2\ln^2\beta_{pi} m_i n_i u_i^2\sim10^{-8}~{\rm dyn}.
\end{displaymath}
Here, we took into account that the ion drift velocity is of the order of the thermal velocity, and estimated the ion number
density $n_i $ from the quasineutrality condition at large Havnes parameter, $n_i = \left| {Z_d } \right|n_d \sim 3\times
10^8$~cm$^{-3}$. Thus, $F_e $ and $F_i$ occur to be of the same order of magnitude. Note that the ion drag force, which
always opposes the electric force, is pointed away from the center.

As regards the interaction between the projectile and dust particles, we assume that the Coulomb interaction is screened at
short distances. Then the upper-bound estimate would suggest total transfer of the momentum from dust particles that find
themselves inside this cylinder to the projectile. The force is given by \cite{Fortov05}
\begin{displaymath}
F_d = \pi \bar R^2 n_d M_d u^2 \sim 10^{-8}~{\rm dyn},
\end{displaymath}
which is not greater than the upper-bound estimates for the plasma-induced forces.

We conclude that the major {\it known} force acting on the projectile is the neutral drag which opposes the {\it unknown}
driving force and, presumably, results in the observed (locally) steady-state motion (the neutral damping rate is
$F_n/M_pu\sim10$~s$^{-1}$, i.e., the transition to steady state occurs within $\sim0.1$~s). Leaving the discussion of
possible driving mechanisms aside, let us focus on the flow of dust particles caused by the projectile. We assume that the
plasma crystal is melted in the neighborhood of the projectile that is streamlined by dust particles. If the stream flow
is a potential nonviscous one the total momentum is conserved, and $F_d $ must vanish. This assumption can be validated by
comparison between dust particle trajectories calculated theoretically and those obtained from experiment.

We apply a hydrodynamic approach to describe the dust particle motion, which is widely used in physics of complex and dusty
plasmas \cite{2}. Based on the discussion above, we can treat the projectile--dust interaction as that of hard spheres
characterized by the length scale $\bar R$. Consider a steady flow of an incompressible liquid of dust particles, described
by the Navier--Stokes equation,
\begin{equation}
({\bf{v}}\cdot\nabla){\bf{v}} = - \nabla \frac{p}{\rho } + \eta\nabla^2{\bf v}, \label{e1}
\end{equation}
where ${\bf{v}}({\bf{r}})$ is the velocity field, $p$ and $\rho$ are the pressure and density, respectively, and $\eta$ is
the shear viscosity of the particle fluid. For an incompressible fluid, when $\rho=$~const, Eq.~(\ref{e1}) is completed by
the continuity equation
\begin{equation}
{\rm div}\:{\bf{v}} = 0. \label{e3}
\end{equation}
Let us estimate the relevant Reynolds number, $\mbox{Re}=\rho u\bar R/\eta$, which characterizes the relative importance of the
viscous term in Eq. (\ref{e1}). Using results of the molecular dynamics simulations of Yukawa fluids \cite{Saigo02}, we
obtain that for our conditions the kinematic viscosity can be estimated as $\eta/\rho=(0.1-0.3)\omega_{pd}\bar r_d^2$, where
$\omega_{pd}=\sqrt{4\pi Z_d^2e^2n_d/M_d}\sim300$~s$^{-1}$ is the dust plasma frequency, so that the Reynolds number is
$\mbox{Re}\sim10$. Thus, we can reasonably neglect the viscosity and thus reduce Eq.~(\ref{e1}) to the Euler equation for an ideal
fluid,
\begin{equation}
\rho ({\bf{v}}\nabla ){\bf{v}} = - \nabla p. \label{e2}
\end{equation}
For an irrotational flow of incompressible liquid (${\rm rot}\:{\bf{v}} = 0$), the substitution ${\bf{v}} = \nabla \varphi $
transforms (\ref{e3}) to the form
\begin{equation}
\nabla^2\varphi = 0, \label{e4}
\end{equation}
where $\varphi $ is the velocity field potential.

Consider a streamline about a sphere with the radius $\bar R$ moving with a constant velocity ${\bf{u}}$. The boundary
conditions are ${\bf{v}}(\infty) = 0$ and ${\bf{v}}(\bar R)\cdot{\bf{n}} = {\bf{u}} \cdot {\bf{n}}$, where ${\bf{r}}$ is the
radius vector from the center of the sphere and ${\bf{n}}={\bf r}/r$ is the radial unit vector. The unique solution of
Eq.~(\ref{e4}) with these boundary conditions is $\varphi({\bf r}) = - (\bar R^3 /2r^2 ){\bf{u}} \cdot {\bf{n}}$, so that
\cite{25}
\begin{equation}
{\bf{v}}({\bf r}) = \frac{{\bar R^3 }}{{2r^3
}}[3{\bf{n}}({\bf{u}} \cdot {\bf{n}}) - {\bf{u}}].
\label{e5}
\end{equation}
The pressure field $p({\bf{r}})$ can be derived from Eq.~(\ref{e2}) but it is unnecessary for our purposes.

The streamlines can be obtained by integration of Eq.~(\ref{e5}) in the laboratory frame of reference. Let the $X$-axis be
directed along the motion of a sphere. Then vector ${\bf{n}}$ has the components $\{ (x - ut)/r,\;y/r,\;z/r\} $, where $r =
\left[ {(x - ut)^2 + y^2 + z^2 } \right]^{1/2} $ and ${\bf{u}} \cdot {\bf{n}} = (x - ut)u/r$. We introduce the variables
$\zeta = (x - ut)/\bar R$, $\eta = y/\bar R$, and $\tau = \nu t$, where $\nu = 3u/2\bar R$, to rewrite Eq.~(\ref{e5}) as a
set of differential equations,
\begin{equation}
\begin{array}{*{20}c}
  {\frac{{d\zeta }}{{d\tau }} = \left( {\frac{{\zeta ^2
}}{{\zeta ^2 + \eta ^2 }} - \frac{1}{3}} \right)\left( {\zeta
^2 + \eta ^2 } \right)^{ - 3/2} - \frac{2}{3},} \\
  {\frac{{d\eta }}{{d\tau }} = \zeta \eta \left( {\zeta ^2 +
\eta ^2 } \right)^{ - 5/2} },
\end{array} \label{e6}
\end{equation}
with the initial conditions $\zeta (0) = 0$ and $\eta (0) = \eta _0 $ ($\left| {\eta _0 } \right| \ge 1$). Due to the
axial symmetry, the third equation for $Z$ -axis coincides with the second equation (\ref{e6}) and is therefore redundant.

Because the decay of particle velocity (\ref{e5}) with the distance from projectile is rather fast, $\left| {\bf{v}} \right|
\propto r^{ - 3} $, we can assume that particles move in a thin fluid tube in the vicinity of a sphere with the radius $\bar
R$ so that the difference between $r$ and $\bar R$ can be neglected and
$\zeta ^2 + \eta ^2 \simeq 1.$
Equations (\ref{e6}) are then reduced to
\begin{equation}
\begin{array}{*{20}c}
  {\frac{{d\zeta }}{{d\tau }} = \zeta ^2 - 1,} \\
  {\frac{{d\eta }}{{d\tau }} = \zeta \eta ,}
\end{array} \label{e7}
\end{equation}
with the initial conditions $\zeta (0) = 0$ and $\eta (0) = 1$. Within the framework of this approximation, we have to
assume also that a particle is quiescent until it finds itself on the surface of a moving sphere (projectile cell). This
instant corresponds to the time $t = - \Delta t$
 and to the distance $d$
 between the particle and direction of projectile motion
(impact parameter). Obviously, solution of Eqs.~(\ref{e6})
obeys the condition $\left| \eta ( \pm \infty )\right|  = d/{\bar R} $. Due to the
time reversibility of Eqs.~(\ref{e7}), the particle must stop
at $t = \Delta t$
 and further remain quiescent, the total time of motion
being $2\Delta t$. Thus, the initial conditions should be
completed by the `collision' condition $\eta (\Delta t) = \pm
d/{\bar R}$, where the sign of $d$
 defines the direction of particle motion. We integrate
Eqs.~(\ref{e7}) to derive $\zeta (\tau ) = - \tanh \tau $
 and $\eta (\tau ) = (\cosh \tau )^{ - 1} $. In conventional
units, we obtain
\begin{equation}
\begin{array}{*{20}c}
  {x^* = \frac{x}{{\bar R}} = \frac{2}{3}\nu t -
\tanh \nu t,} \\
  {y^* = \frac{y}{{\bar R}} = \frac{1}{{\cosh
\nu t}}.} \\
\end{array} \label{e8}
\end{equation}
From the second equation (\ref{e8}) we obtain
\begin{equation}
\Delta t = \nu ^{ - 1} \ln \left( {\frac{{\bar R}}{d} + \sqrt
{\frac{{\bar R^2 }}{{d^2 }} - 1} }\;\right). \label{e9}
\end{equation}
If $1 - (d/\bar R) \ll 1$, we find approximately $\Delta t \simeq \left( {2\bar R/d - 2} \right)^{1/2} $ for a grazing
collision. At $(d/\bar R) \ll 1$, we have $\Delta t \simeq \ln (2\bar R/d)$. Streamline (\ref{e8}) is shown in Fig.~\ref{f4}
for $d/\bar R = 0.163$, which corresponds to $\nu \Delta t = 2.5$ in the framework of used approximation. It is seen in
Fig.~\ref{f4} that the approximate solution almost coincides with the numerical solution of equation set (\ref{e6})
corresponding to the same $d/\bar R$. Our approximation is valid even at a large impact parameter $d = \bar R$ because in
this case, numerical solution of (\ref{e6}) yields $y(0) = 1.33\bar R$, i.e., the distance between the particle and
projectile is still not much different from $\bar R$. Figure~\ref{f5} shows a good correspondence between solution
(\ref{e8}), (\ref{e9}) and experimentally observed traces of individual particles extracted from Fig.~\ref{f3}c in their
closed portions. Indication of this correspondence is a good agreement between the ratio of the height of a closed loop to
$\bar R$ determined experimentally (Figs.~\ref{f3}c and \ref{f5}), which amounts to about $0.28$, while the theoretical
value is $0.277$ (cf. Fig.~\ref{f4}). This ratio is independent of the loop position relative to the projectile path.
Note that since the streamline takes place in a close neighborhood of the projectile and farther regions are crystallized,
an approximate solution (\ref{e8}), (\ref{e9}) utilizing the same approximation seems to be even more adequate to treated
system than the numerical solution. Open portions of particle traces reveal some differences from theory. In Fig.~\ref{f5},
the upper trace portion of the bottom loop on the left-hand side seems to disappear due to the fact that the particle go out
of the illumination plane. On the contrary, the upper trace portion of the top loop on the left-hand side would not appear
at all but the particle was probably pushed towards the projectile due to spatial re-distribution of particles in the course
of local crystallization.

We have demonstrated that the stream flow of dust particles is almost a potential nonviscous one. In an ideal model, the
momentum transferred from dust particles to projectile and back must coincide, so that the force $F_d $ of interaction
between the dust particles and projectile vanishes. In reality, however, $F_d $ is never equal to zero. It follows from
solutions for the particle trajectories obtained above that the motion of particles {\it outside} the cylinder (whose axis
coincides with the projectile trajectory) can be disregarded, as we originally assumed. The lower-bound estimate stems from the momentum not
transferred back to the projectile, due to collisions between neutrals and dust particles moving along the loops with the
velocity estimated as $u/2$:
\[ F_d \simeq \frac{{F_n }}{2}\left( {\frac{{a_d }}{{a_p
}}} \right)^2 \frac{{4\pi }}{3}n_d \bar R^3 \sim 10^{-8} \;{\mbox{dyn}}{\mbox{,}} \] 
where $a_d = 1.275 \times 10^{ -4}$~cm is the dust particle radius.
Thus, $F_d$ proves to be of the same
order of magnitude as its upper-bound estimate obtained above and as $F_n$. However, if treated damping effect was taken into
account in Eq.~(\ref{e1}) the solution for trajectories (\ref{e8}), (\ref{e9}) would not change because it is a consequence
of unchanged Eq.~(\ref{e3}) with the same boundary conditions.

\section{\label{s4} CONCLUSION}

In this study, we have observed that a heavy particle (projectile) going through the dust crystal with a relatively low
(subsonic) velocity appears to move almost freely. We have proposed a new explanation of this fact implying a nonviscous
flow of the dust particles about the projectile,
in which the force of interaction between the projectile and dust particles vanishes. To prove this, we monitored the
particle dynamics by means of a high-resolution camera. Superposition of successive frames revealed typical $\alpha$-shaped
pathways of dust particles in the neighborhood of projectile. Since the system is strongly coupled, it can be divided into
the Wigner--Seitz cells around each particle including the projectile, whose motion does not break close ordering. However,
the projectile melts the crystal in its neighborhood, and the dust particles flow about the projectile cell boundary in such
a way that the distance $\bar R$ between individual dust particles and the projectile, which can be measured using recorded
frames, is the sum of radii of the Wigner--Seitz cells around dust particles and the projectile. In the framework of our
model, we have to imply that the cells around particles rather than the particles themselves take part in the interaction.
In the absence of direct interparticle contacts, the shear viscosity must vanish and we can consider the nonviscous flow of
the cells.

The classical solution for the velocity field in a potential
nonviscous flow about a sphere moving with a constant
velocity can be regarded as a set of differential equations
for the trajectories of individual dust particles. We have
solved these equations numerically and obtained an
analytical solution, which approximates closely the
numerical one. This solution, which defines the coordinates
of dust particles as explicit functions of time, is in a good
agreement with particle trajectories obtained experimentally
by superposition of successive frames. This supports our
main assumption concerning the nonviscous nature of
particle flow. Because such motion implies no momentum
transfer from liquid to the projectile, the drag force $F_d $
 arising from the interaction between the projectile and dust
particles must {\it vanish\/}. Note that examples of such
type of dust particle collective motion were previously not
encountered in physics of complex and dusty plasmas.

\begin{figure}
\includegraphics[width=6.6in]{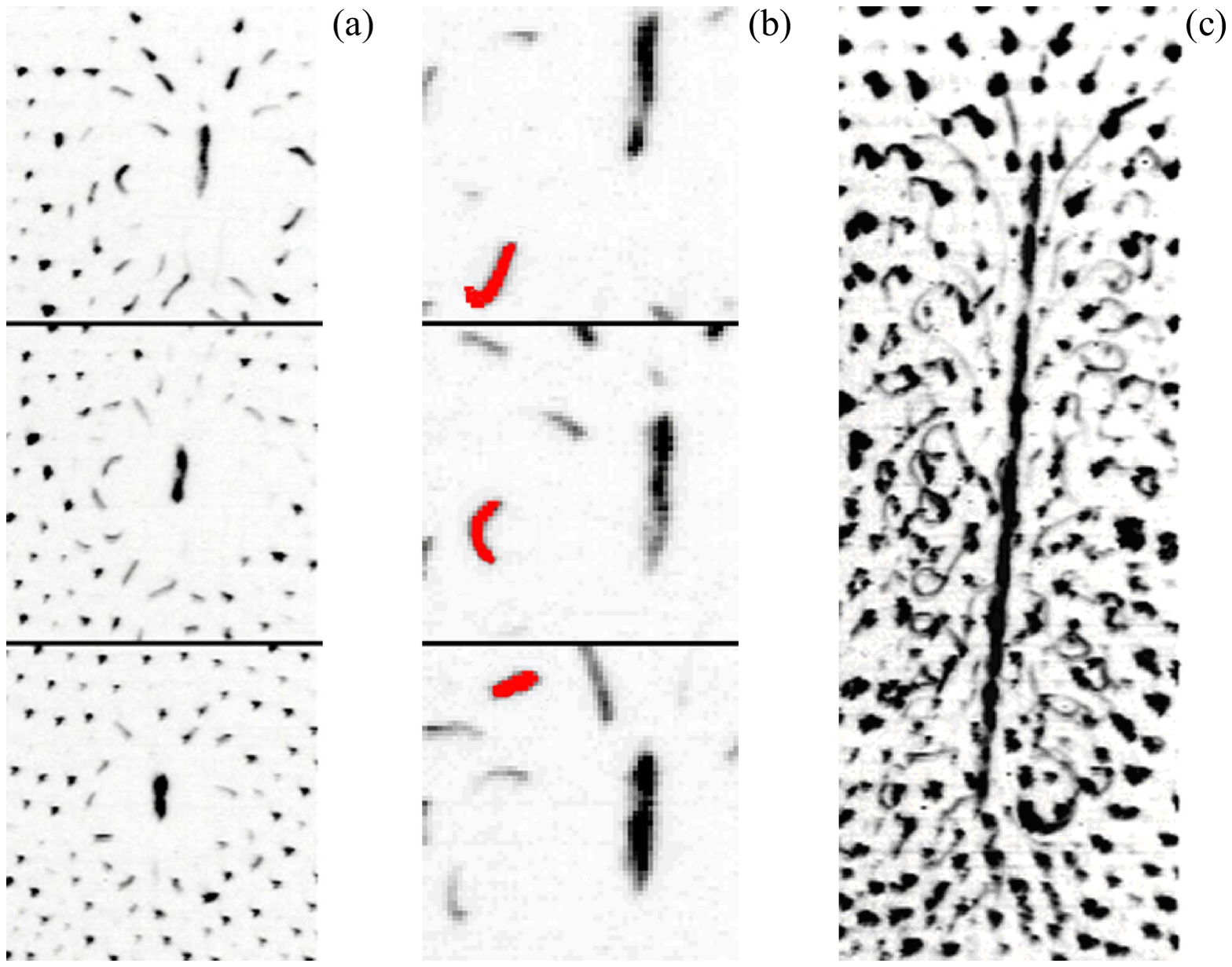}
\caption{\label{f3}}
\end{figure}

A real system does not show idealized behavior. Not all the particles move along the $\alpha$-shaped pathways, which we
attribute (i) to the fact that the illumination plane does not coincides with one of the crystalline planes and is not
exactly parallel to the direction of the projectile motion, (ii) to the finite thickness of the laser sheet, so that
particles can leave the illuminated space, and (iii) to a local crystallization after the projectile passage. In addition,
the force of interaction between the projectile and dust particles does not vanish exactly. However, a general conclusion of
the developed theory, that only the dust particles inside a cylinder with the radius $\bar R$ are involved in collective
motion, is strongly confirmed by Fig.~\ref{f3}a. Note that this is a consequence of the boundary condition imposed on the
projectile cell, which implies a nonzero tangential velocity field component. In the case of a viscous flow (e.g.,
Refs~\cite{23,Ivlev07}) the situation can be different due to vanishing of this component, resulting to much larger number
of moving particles.

We can conclude that a large particle can move almost
freely inside the bulk of a strongly coupled dust crystal. Our
investigation also points to the fact that a hydrodynamic
approach can be valid at small length scales down to
several interparticle distances. The processes of local dust
crystal melting and freezing, which were not treated in
detail in the present study, will be addressed in future work.

\begin{acknowledgments}
The microgravity research is supported by the space agency of the 
Deutsches Zentrum f\"ur Luft- und Raumfahrt e.V., with funds from the 
federalministry for economy and technology according to a resolution of 
the Deutscher Bundestag under grant No. 50WP0203 and 50WM1203. We 
additionally acknowledge financial support from the European Research 
Council under the European Union's Seventh Framework Programme 
(FP7/2007-2013) / ERC Grant agreement 267499, and from the Russian Basic 
Research Foundation and ROSCOSMOS. We like to thank the industry teams 
from RSC-Energia and Kayser-Threde and the cosmonauts who turned the 
ambitious research on the ISS into a success.
\end{acknowledgments}

\clearpage
\providecommand{\noopsort}[1]{}\providecommand{\singleletter}[1]{#1}%

\newpage

\centerline{FIGURE CAPTIONS}
\vskip\baselineskip

\begin{figure}
\includegraphics[width=6.6in]{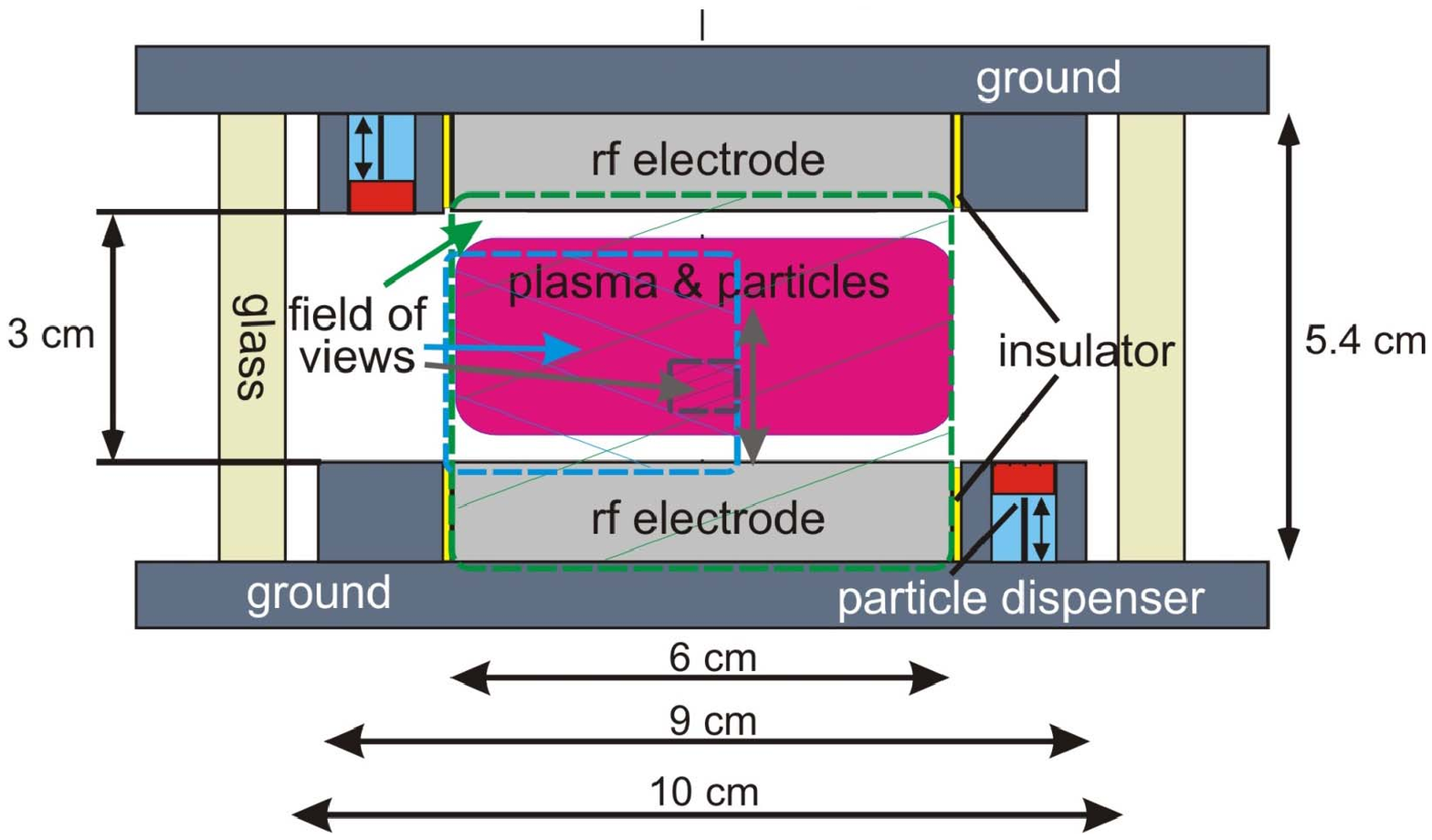}
\caption{\label{f1}}
\end{figure}

Fig.~\ref{f1}: The cross-sectional schematic of the plasma
chamber
\vskip\baselineskip

\begin{figure}
\includegraphics[width=6.6in]{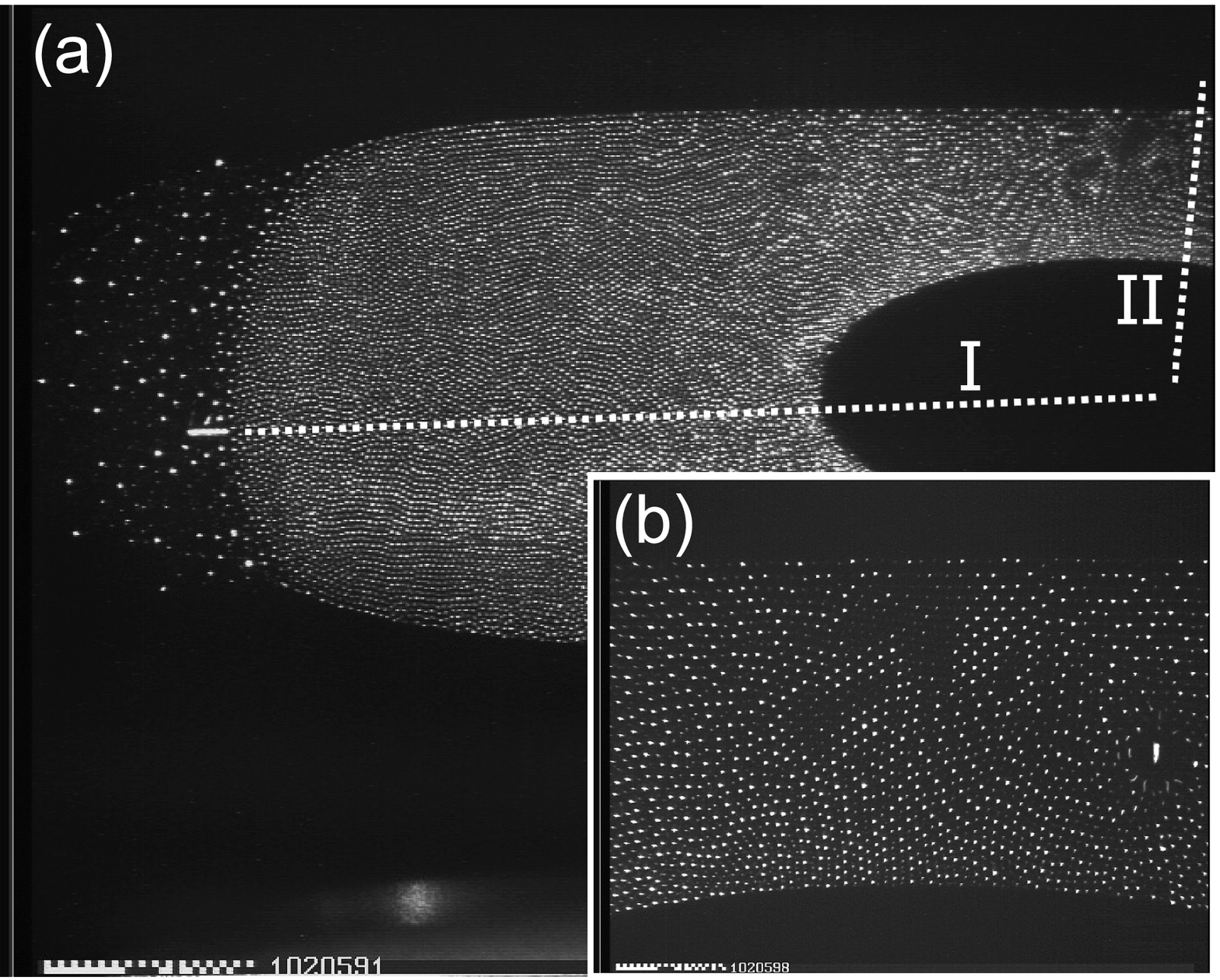}
\caption{\label{f2}}
\end{figure}

Fig.~\ref{f2}: (a), snapshot of the microparticle cloud from
quadrant view camera (field of view is 35.7 mm x
26.0 mm). On the left-hand side, a small track of initial
movement of the probe particle (projectile) into the cloud is
visible. The horizontal dashed line arrow shows
schematically the path of projectile on the left-hand side
(phase I); the vertical dashed line arrow, the path of
projectile from the void into the upper part of microparticle
cloud (phase II). (b), snapshot of the microparticle cloud
from high-resolution view camera (field of view is 8.1 mm
x 5.9 mm). A small track on the right-hand side shows the
projectile motion in phase II
\vskip\baselineskip

Fig.~\ref{f3}: Recorded positions of dust particles and the
projectile (negative images). Time interval between
individual frames is (a), 0.1 s and (b), 0.02 s; (c) is a
superposition of 12 frames with the time interval of 0.02 s.
The projectile motion from void to chamber wall
corresponds to frames from bottom to top
\vskip\baselineskip

\begin{figure}
\includegraphics[width=6.6in]{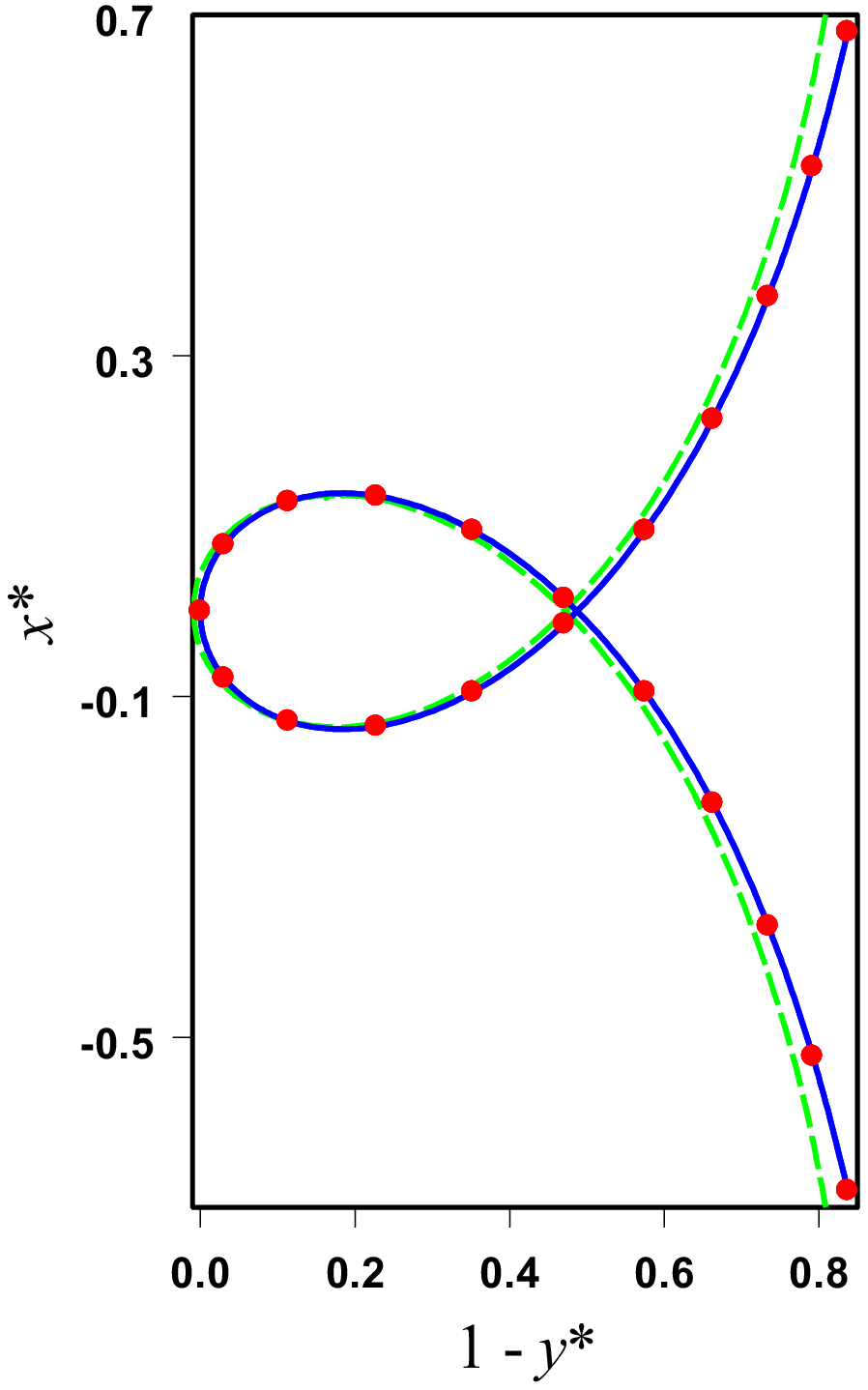}
\caption{\label{f4}}
\end{figure}

Fig.~\ref{f4}: Calculated trajectory of a dust particle
moving about the projectile for $\nu \Delta t = 2.5$.
Solid line shows calculation using (\ref{e8}), (\ref{e9}) and
dashed line, numerical integration of Eqs.~(\ref{e6}) for the
same $d = 0.163{\bar R}$. Dots mark the intervals $\nu t = 1/4$

\vskip\baselineskip

\begin{figure}
\includegraphics[width=6.6in]{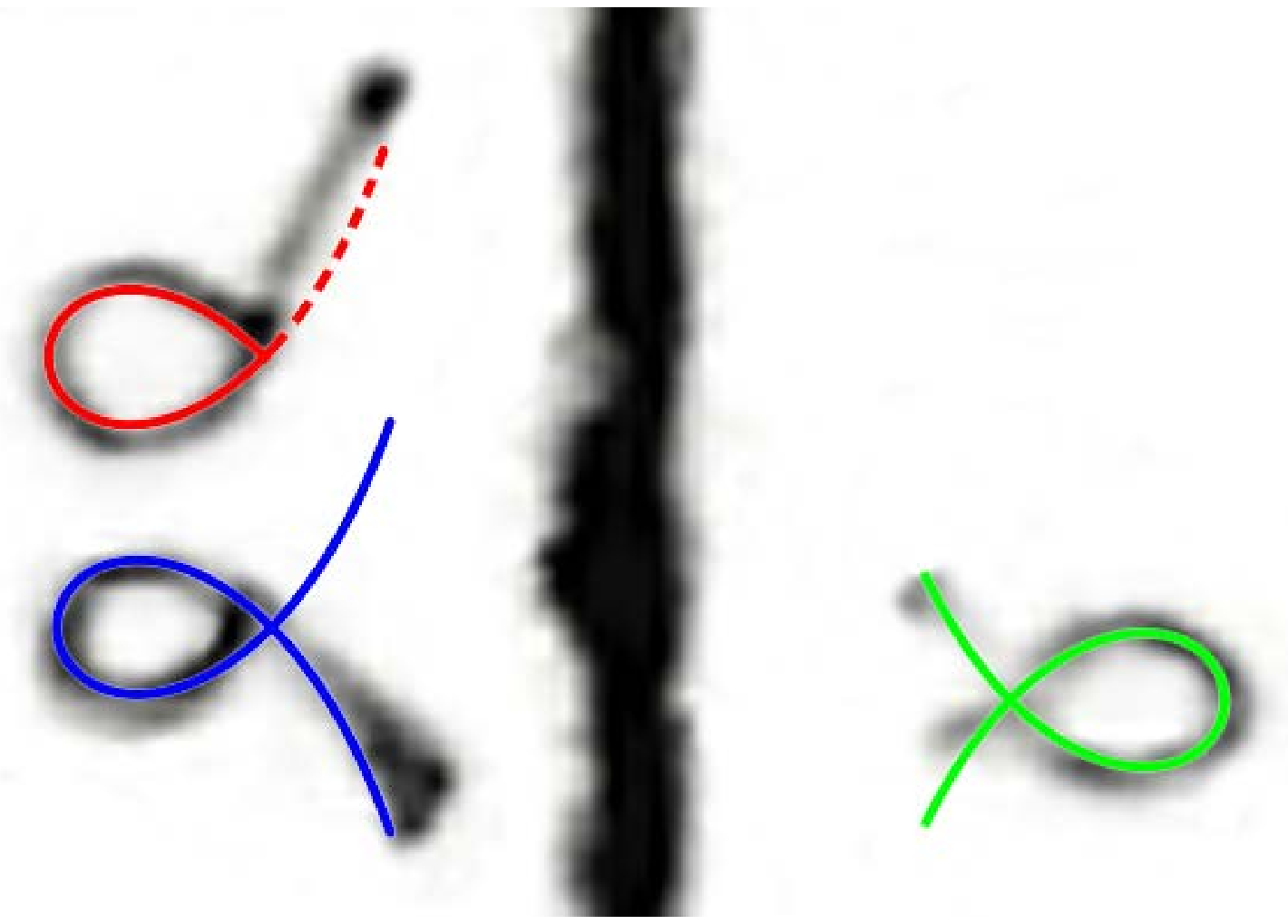}
\caption{\label{f5}}
\end{figure}

Fig.~\ref{f5}: Individual dust particles moving along the
$\alpha$-shaped pathways. Traces are the superposition of
frames obtained in the experiment (negative
image, vertical trace marks the projectile trajectory), solid
lines show calculation by formulas (\ref{e8}) and
(\ref{e9}), dashed line indicates the trajectory portion
`forced' by local crystallization

\newpage
\pagestyle{empty}

\newpage
\pagestyle{empty}

\newpage
\pagestyle{empty}

\newpage
\pagestyle{empty}

\newpage
\pagestyle{empty}


\begin{thebibliography}{36}%
\makeatletter
\providecommand \@ifxundefined [1]{%
 \@ifx{#1\undefined}
}%
\providecommand \@ifnum [1]{%
 \ifnum #1\expandafter \@firstoftwo
 \else \expandafter \@secondoftwo
 \fi
}%
\providecommand \@ifx [1]{%
 \ifx #1\expandafter \@firstoftwo
 \else \expandafter \@secondoftwo
 \fi
}%
\providecommand \natexlab [1]{#1}%
\providecommand \enquote  [1]{``#1''}%
\providecommand \bibnamefont  [1]{#1}%
\providecommand \bibfnamefont [1]{#1}%
\providecommand \citenamefont [1]{#1}%
\providecommand \href@noop [0]{\@secondoftwo}%
\providecommand \href [0]{\begingroup \@sanitize@url \@href}%
\providecommand \@href[1]{\@@startlink{#1}\@@href}%
\providecommand \@@href[1]{\endgroup#1\@@endlink}%
\providecommand \@sanitize@url [0]{\catcode `\\12\catcode `\$12\catcode
  `\&12\catcode `\#12\catcode `\^12\catcode `\_12\catcode `\%12\relax}%
\providecommand \@@startlink[1]{}%
\providecommand \@@endlink[0]{}%
\providecommand \url  [0]{\begingroup\@sanitize@url \@url }%
\providecommand \@url [1]{\endgroup\@href {#1}{\urlprefix }}%
\providecommand \urlprefix  [0]{URL }%
\providecommand \Eprint [0]{\href }%
\providecommand \doibase [0]{http://dx.doi.org/}%
\providecommand \selectlanguage [0]{\@gobble}%
\providecommand \bibinfo  [0]{\@secondoftwo}%
\providecommand \bibfield  [0]{\@secondoftwo}%
\providecommand \translation [1]{[#1]}%
\providecommand \BibitemOpen [0]{}%
\providecommand \bibitemStop [0]{}%
\providecommand \bibitemNoStop [0]{.\EOS\space}%
\providecommand \EOS [0]{\spacefactor3000\relax}%
\providecommand \BibitemShut  [1]{\csname bibitem#1\endcsname}%
\let\auto@bib@innerbib\@empty
\bibitem [{\citenamefont {Morfill}\ and\ \citenamefont
  {Ivlev}(2009)}]{Morfill09}%
  \BibitemOpen
  \bibfield  {author} {\bibinfo {author} {\bibfnamefont {G.~E.}\ \bibnamefont
  {Morfill}}\ and\ \bibinfo {author} {\bibfnamefont {A.~V.}\ \bibnamefont
  {Ivlev}},\ }\href@noop {} {\bibfield  {journal} {\bibinfo  {journal} {Rev.\
  Mod.\ Phys.}\ }\textbf {\bibinfo {volume} {81}},\ \bibinfo {pages} {1353}
  (\bibinfo {year} {2009})}\BibitemShut {NoStop}%
\bibitem [{\citenamefont {Fortov}\ and\ \citenamefont {Morfill}(2009)}]{2}%
  \BibitemOpen
  \bibinfo {editor} {\bibfnamefont {V.~E.}\ \bibnamefont {Fortov}}\ and\
  \bibinfo {editor} {\bibfnamefont {G.~E.}\ \bibnamefont {Morfill}},\ eds.,\
  \href@noop {} {\emph {\bibinfo {title} {Complex and Dusty Plasmas: From
  Laboratory to Space}}},\ Series in Plasma Physics\ (\bibinfo  {publisher}
  {CRC Press, New York},\ \bibinfo {year} {2009})\BibitemShut {NoStop}%
\bibitem [{\citenamefont {Thomas}\ and\ \citenamefont {Morfill}(1996)}]{1}%
  \BibitemOpen
  \bibfield  {author} {\bibinfo {author} {\bibfnamefont {H.~M.}\ \bibnamefont
  {Thomas}}\ and\ \bibinfo {author} {\bibfnamefont {G.~E.}\ \bibnamefont
  {Morfill}},\ }\href@noop {} {\bibfield  {journal} {\bibinfo  {journal}
  {Nature}\ }\textbf {\bibinfo {volume} {379}},\ \bibinfo {pages} {806}
  (\bibinfo {year} {1996})}\BibitemShut {NoStop}%
\bibitem [{\citenamefont {Chu}\ and\ \citenamefont {I}(1994)}]{3}%
  \BibitemOpen
  \bibfield  {author} {\bibinfo {author} {\bibfnamefont {J.~H.}\ \bibnamefont
  {Chu}}\ and\ \bibinfo {author} {\bibfnamefont {L.}~\bibnamefont {I}},\
  }\href@noop {} {\bibfield  {journal} {\bibinfo  {journal} {Phys.\ Rev.\
  Lett.}\ }\textbf {\bibinfo {volume} {72}},\ \bibinfo {pages} {4009} (\bibinfo
  {year} {1994})}\BibitemShut {NoStop}%
\bibitem [{\citenamefont {Thomas}\ \emph {et~al.}(1994)\citenamefont {Thomas},
  \citenamefont {Morfill}, \citenamefont {Demmel}, \citenamefont {Goree},
  \citenamefont {Feuerbacher},\ and\ \citenamefont {M{\"{o}}hlmann}}]{4}%
  \BibitemOpen
  \bibfield  {author} {\bibinfo {author} {\bibfnamefont {H.}~\bibnamefont
  {Thomas}}, \bibinfo {author} {\bibfnamefont {G.~E.}\ \bibnamefont {Morfill}},
  \bibinfo {author} {\bibfnamefont {V.}~\bibnamefont {Demmel}}, \bibinfo
  {author} {\bibfnamefont {J.}~\bibnamefont {Goree}}, \bibinfo {author}
  {\bibfnamefont {B.}~\bibnamefont {Feuerbacher}}, \ and\ \bibinfo {author}
  {\bibfnamefont {D.}~\bibnamefont {M{\"{o}}hlmann}},\ }\href@noop {}
  {\bibfield  {journal} {\bibinfo  {journal} {Phys.\ Rev.\ Lett.}\ }\textbf
  {\bibinfo {volume} {73}},\ \bibinfo {pages} {652} (\bibinfo {year}
  {1994})}\BibitemShut {NoStop}%
\bibitem [{\citenamefont {Melzer}\ \emph {et~al.}(1994)\citenamefont {Melzer},
  \citenamefont {Trottenberg},\ and\ \citenamefont {Piel}}]{5}%
  \BibitemOpen
  \bibfield  {author} {\bibinfo {author} {\bibfnamefont {A.}~\bibnamefont
  {Melzer}}, \bibinfo {author} {\bibfnamefont {T.}~\bibnamefont {Trottenberg}},
  \ and\ \bibinfo {author} {\bibfnamefont {A.}~\bibnamefont {Piel}},\
  }\href@noop {} {\bibfield  {journal} {\bibinfo  {journal} {Phys.\ Lett.}\
  }\textbf {\bibinfo {volume} {191}},\ \bibinfo {pages} {301} (\bibinfo {year}
  {1994})}\BibitemShut {NoStop}%
\bibitem [{\citenamefont {Hayashi}(1999)}]{6}%
  \BibitemOpen
  \bibfield  {author} {\bibinfo {author} {\bibfnamefont {Y.}~\bibnamefont
  {Hayashi}},\ }\href@noop {} {\bibfield  {journal} {\bibinfo  {journal}
  {Phys.\ Rev.\ Lett.}\ }\textbf {\bibinfo {volume} {83}},\ \bibinfo {pages}
  {4764} (\bibinfo {year} {1999})}\BibitemShut {NoStop}%
\bibitem [{\citenamefont {Fortov}\ \emph {et~al.}(2005)\citenamefont {Fortov},
  \citenamefont {Ivlev}, \citenamefont {Khrapak}, \citenamefont {Khrapak},\
  and\ \citenamefont {Morfill}}]{Fortov05}%
  \BibitemOpen
  \bibfield  {author} {\bibinfo {author} {\bibfnamefont {V.}~\bibnamefont
  {Fortov}}, \bibinfo {author} {\bibfnamefont {A.}~\bibnamefont {Ivlev}},
  \bibinfo {author} {\bibfnamefont {S.}~\bibnamefont {Khrapak}}, \bibinfo
  {author} {\bibfnamefont {A.}~\bibnamefont {Khrapak}}, \ and\ \bibinfo
  {author} {\bibfnamefont {G.}~\bibnamefont {Morfill}},\ }\href@noop {}
  {\bibfield  {journal} {\bibinfo  {journal} {Phys.\ Rep.}\ }\textbf {\bibinfo
  {volume} {421}},\ \bibinfo {pages} {1} (\bibinfo {year} {2005})}\BibitemShut
  {NoStop}%
\bibitem [{\citenamefont {Fortov}\ \emph
  {et~al.}(2004{\natexlab{a}})\citenamefont {Fortov}, \citenamefont {Khrapak},
  \citenamefont {Khrapak}, \citenamefont {Molotkov},\ and\ \citenamefont
  {Petrov}}]{7}%
  \BibitemOpen
  \bibfield  {author} {\bibinfo {author} {\bibfnamefont {V.~E.}\ \bibnamefont
  {Fortov}}, \bibinfo {author} {\bibfnamefont {A.~G.}\ \bibnamefont {Khrapak}},
  \bibinfo {author} {\bibfnamefont {S.~A.}\ \bibnamefont {Khrapak}}, \bibinfo
  {author} {\bibfnamefont {V.~I.}\ \bibnamefont {Molotkov}}, \ and\ \bibinfo
  {author} {\bibfnamefont {O.~F.}\ \bibnamefont {Petrov}},\ }\href@noop {}
  {\bibfield  {journal} {\bibinfo  {journal} {Phys.-Usp.}\ }\textbf {\bibinfo
  {volume} {47}},\ \bibinfo {pages} {447} (\bibinfo {year}
  {2004}{\natexlab{a}})}\BibitemShut {NoStop}%
\bibitem [{\citenamefont {Fortov}\ \emph {et~al.}(1996)\citenamefont {Fortov},
  \citenamefont {Nefedov}, \citenamefont {Torchinskii}, \citenamefont
  {Molotkov}, \citenamefont {Khrapak}, \citenamefont {Petrov},\ and\
  \citenamefont {Volykhin}}]{8}%
  \BibitemOpen
  \bibfield  {author} {\bibinfo {author} {\bibfnamefont {V.~E.}\ \bibnamefont
  {Fortov}}, \bibinfo {author} {\bibfnamefont {A.~P.}\ \bibnamefont {Nefedov}},
  \bibinfo {author} {\bibfnamefont {V.~M.}\ \bibnamefont {Torchinskii}},
  \bibinfo {author} {\bibfnamefont {V.~I.}\ \bibnamefont {Molotkov}}, \bibinfo
  {author} {\bibfnamefont {A.~G.}\ \bibnamefont {Khrapak}}, \bibinfo {author}
  {\bibfnamefont {O.~F.}\ \bibnamefont {Petrov}}, \ and\ \bibinfo {author}
  {\bibfnamefont {K.~F.}\ \bibnamefont {Volykhin}},\ }\href@noop {} {\bibfield
  {journal} {\bibinfo  {journal} {JETP Lett.}\ }\textbf {\bibinfo {volume}
  {64}},\ \bibinfo {pages} {92} (\bibinfo {year} {1996})}\BibitemShut {NoStop}%
\bibitem [{\citenamefont {{\relax Thomas, Jr.}}\ \emph
  {et~al.}(2007)\citenamefont {{\relax Thomas, Jr.}}, \citenamefont {Fisher},\
  and\ \citenamefont {Merlino}}]{EThomas}%
  \BibitemOpen
  \bibfield  {author} {\bibinfo {author} {\bibfnamefont {E.}~\bibnamefont
  {{\relax Thomas, Jr.}}}, \bibinfo {author} {\bibfnamefont {R.}~\bibnamefont
  {Fisher}}, \ and\ \bibinfo {author} {\bibfnamefont {R.~L.}\ \bibnamefont
  {Merlino}},\ }\href@noop {} {\bibfield  {journal} {\bibinfo  {journal}
  {Phys.\ Plasmas}\ }\textbf {\bibinfo {volume} {14}},\ \bibinfo {pages}
  {123701} (\bibinfo {year} {2007})}\BibitemShut {NoStop}%
\bibitem [{\citenamefont {Morfill}\ \emph {et~al.}(2006)\citenamefont
  {Morfill}, \citenamefont {Konopka}, \citenamefont {Kretschmer}, \citenamefont
  {Rubin-Zuzic}, \citenamefont {Thomas}, \citenamefont {Zhdanov},\ and\
  \citenamefont {Tsytovich}}]{9}%
  \BibitemOpen
  \bibfield  {author} {\bibinfo {author} {\bibfnamefont {G.~E.}\ \bibnamefont
  {Morfill}}, \bibinfo {author} {\bibfnamefont {U.}~\bibnamefont {Konopka}},
  \bibinfo {author} {\bibfnamefont {M.}~\bibnamefont {Kretschmer}}, \bibinfo
  {author} {\bibfnamefont {M.}~\bibnamefont {Rubin-Zuzic}}, \bibinfo {author}
  {\bibfnamefont {H.~M.}\ \bibnamefont {Thomas}}, \bibinfo {author}
  {\bibfnamefont {S.~K.}\ \bibnamefont {Zhdanov}}, \ and\ \bibinfo {author}
  {\bibfnamefont {V.}~\bibnamefont {Tsytovich}},\ }\href@noop {} {\bibfield
  {journal} {\bibinfo  {journal} {New J.\ Phys.}\ }\textbf {\bibinfo {volume}
  {8}},\ \bibinfo {pages} {7} (\bibinfo {year} {2006})}\BibitemShut {NoStop}%
\bibitem [{\citenamefont {Caliebe}\ \emph {et~al.}(2011)\citenamefont
  {Caliebe}, \citenamefont {Arp},\ and\ \citenamefont {Piel}}]{10}%
  \BibitemOpen
  \bibfield  {author} {\bibinfo {author} {\bibfnamefont {D.}~\bibnamefont
  {Caliebe}}, \bibinfo {author} {\bibfnamefont {O.}~\bibnamefont {Arp}}, \ and\
  \bibinfo {author} {\bibfnamefont {A.}~\bibnamefont {Piel}},\ }\href@noop {}
  {\bibfield  {journal} {\bibinfo  {journal} {Phys.\ of Plasmas}\ }\textbf
  {\bibinfo {volume} {18}},\ \bibinfo {pages} {073702} (\bibinfo {year}
  {2011})}\BibitemShut {NoStop}%
\bibitem [{\citenamefont {Piel}\ \emph {et~al.}(2008)\citenamefont {Piel},
  \citenamefont {Arp}, \citenamefont {Klindworth},\ and\ \citenamefont
  {Melzer}}]{11}%
  \BibitemOpen
  \bibfield  {author} {\bibinfo {author} {\bibfnamefont {A.}~\bibnamefont
  {Piel}}, \bibinfo {author} {\bibfnamefont {O.}~\bibnamefont {Arp}}, \bibinfo
  {author} {\bibfnamefont {M.}~\bibnamefont {Klindworth}}, \ and\ \bibinfo
  {author} {\bibfnamefont {A.}~\bibnamefont {Melzer}},\ }\href@noop {}
  {\bibfield  {journal} {\bibinfo  {journal} {Phys.\ Rev.\ E}\ }\textbf
  {\bibinfo {volume} {77}},\ \bibinfo {pages} {026407} (\bibinfo {year}
  {2008})}\BibitemShut {NoStop}%
\bibitem [{\citenamefont {Menzel}\ \emph {et~al.}(2011)\citenamefont {Menzel},
  \citenamefont {Arp},\ and\ \citenamefont {Piel}}]{12}%
  \BibitemOpen
  \bibfield  {author} {\bibinfo {author} {\bibfnamefont {K.~O.}\ \bibnamefont
  {Menzel}}, \bibinfo {author} {\bibfnamefont {O.}~\bibnamefont {Arp}}, \ and\
  \bibinfo {author} {\bibfnamefont {A.}~\bibnamefont {Piel}},\ }\href@noop {}
  {\bibfield  {journal} {\bibinfo  {journal} {Phys.\ Rev.\ E}\ }\textbf
  {\bibinfo {volume} {83}},\ \bibinfo {pages} {016402} (\bibinfo {year}
  {2011})}\BibitemShut {NoStop}%
\bibitem [{\citenamefont {Arp}\ \emph {et~al.}(2011)\citenamefont {Arp},
  \citenamefont {Caliebe},\ and\ \citenamefont {Piel}}]{13}%
  \BibitemOpen
  \bibfield  {author} {\bibinfo {author} {\bibfnamefont {O.}~\bibnamefont
  {Arp}}, \bibinfo {author} {\bibfnamefont {D.}~\bibnamefont {Caliebe}}, \ and\
  \bibinfo {author} {\bibfnamefont {A.}~\bibnamefont {Piel}},\ }\href@noop {}
  {\bibfield  {journal} {\bibinfo  {journal} {Phys.\ Rev.\ E}\ }\textbf
  {\bibinfo {volume} {83}},\ \bibinfo {pages} {066404} (\bibinfo {year}
  {2011})}\BibitemShut {NoStop}%
\bibitem [{\citenamefont {Schwabe}\ \emph {et~al.}(2008)\citenamefont
  {Schwabe}, \citenamefont {Zhdanov}, \citenamefont {Thomas}, \citenamefont
  {Ivlev}, \citenamefont {Rubin-Zuzic}, \citenamefont {Morfill}, \citenamefont
  {Molotkov}, \citenamefont {Lipaev}, \citenamefont {Fortov},\ and\
  \citenamefont {Reiter}}]{14}%
  \BibitemOpen
  \bibfield  {author} {\bibinfo {author} {\bibfnamefont {M.}~\bibnamefont
  {Schwabe}}, \bibinfo {author} {\bibfnamefont {S.~K.}\ \bibnamefont
  {Zhdanov}}, \bibinfo {author} {\bibfnamefont {H.~M.}\ \bibnamefont {Thomas}},
  \bibinfo {author} {\bibfnamefont {A.~V.}\ \bibnamefont {Ivlev}}, \bibinfo
  {author} {\bibfnamefont {M.}~\bibnamefont {Rubin-Zuzic}}, \bibinfo {author}
  {\bibfnamefont {G.~E.}\ \bibnamefont {Morfill}}, \bibinfo {author}
  {\bibfnamefont {V.~I.}\ \bibnamefont {Molotkov}}, \bibinfo {author}
  {\bibfnamefont {A.~M.}\ \bibnamefont {Lipaev}}, \bibinfo {author}
  {\bibfnamefont {V.~E.}\ \bibnamefont {Fortov}}, \ and\ \bibinfo {author}
  {\bibfnamefont {T.}~\bibnamefont {Reiter}},\ }\href@noop {} {\bibfield
  {journal} {\bibinfo  {journal} {New J.\ Phys.}\ }\textbf {\bibinfo {volume}
  {10}},\ \bibinfo {pages} {033037} (\bibinfo {year} {2008})}\BibitemShut
  {NoStop}%
\bibitem [{\citenamefont {Morfill}\ \emph {et~al.}(1999)\citenamefont
  {Morfill}, \citenamefont {Thomas}, \citenamefont {Konopka}, \citenamefont
  {Rothermel}, \citenamefont {Zuzic}, \citenamefont {Ivlev},\ and\
  \citenamefont {Goree}}]{15}%
  \BibitemOpen
  \bibfield  {author} {\bibinfo {author} {\bibfnamefont {G.~E.}\ \bibnamefont
  {Morfill}}, \bibinfo {author} {\bibfnamefont {H.~M.}\ \bibnamefont {Thomas}},
  \bibinfo {author} {\bibfnamefont {U.}~\bibnamefont {Konopka}}, \bibinfo
  {author} {\bibfnamefont {H.}~\bibnamefont {Rothermel}}, \bibinfo {author}
  {\bibfnamefont {M.}~\bibnamefont {Zuzic}}, \bibinfo {author} {\bibfnamefont
  {A.}~\bibnamefont {Ivlev}}, \ and\ \bibinfo {author} {\bibfnamefont
  {J.}~\bibnamefont {Goree}},\ }\href@noop {} {\bibfield  {journal} {\bibinfo
  {journal} {Phys.\ Rev.\ Lett.}\ }\textbf {\bibinfo {volume} {83}},\ \bibinfo
  {pages} {1598} (\bibinfo {year} {1999})}\BibitemShut {NoStop}%
\bibitem [{\citenamefont {Khrapak}\ \emph {et~al.}(2011)\citenamefont
  {Khrapak}, \citenamefont {Klumov}, \citenamefont {Huber}, \citenamefont
  {Molotkov}, \citenamefont {Lipaev}, \citenamefont {Naumkin}, \citenamefont
  {Thomas}, \citenamefont {Ivlev}, \citenamefont {Morfill}, \citenamefont
  {Petrov}, \citenamefont {Fortov}, \citenamefont {Malentschenko},\ and\
  \citenamefont {Volkov}}]{16}%
  \BibitemOpen
  \bibfield  {author} {\bibinfo {author} {\bibfnamefont {S.~A.}\ \bibnamefont
  {Khrapak}}, \bibinfo {author} {\bibfnamefont {B.~A.}\ \bibnamefont {Klumov}},
  \bibinfo {author} {\bibfnamefont {P.}~\bibnamefont {Huber}}, \bibinfo
  {author} {\bibfnamefont {V.~I.}\ \bibnamefont {Molotkov}}, \bibinfo {author}
  {\bibfnamefont {A.~M.}\ \bibnamefont {Lipaev}}, \bibinfo {author}
  {\bibfnamefont {V.~N.}\ \bibnamefont {Naumkin}}, \bibinfo {author}
  {\bibfnamefont {H.~M.}\ \bibnamefont {Thomas}}, \bibinfo {author}
  {\bibfnamefont {A.~V.}\ \bibnamefont {Ivlev}}, \bibinfo {author}
  {\bibfnamefont {G.~E.}\ \bibnamefont {Morfill}}, \bibinfo {author}
  {\bibfnamefont {O.~F.}\ \bibnamefont {Petrov}}, \bibinfo {author}
  {\bibfnamefont {V.~E.}\ \bibnamefont {Fortov}}, \bibinfo {author}
  {\bibfnamefont {Y.}~\bibnamefont {Malentschenko}}, \ and\ \bibinfo {author}
  {\bibfnamefont {S.}~\bibnamefont {Volkov}},\ }\href@noop {} {\bibfield
  {journal} {\bibinfo  {journal} {Phys.\ Rev.\ Lett.}\ }\textbf {\bibinfo
  {volume} {106}},\ \bibinfo {pages} {205001} (\bibinfo {year}
  {2011})}\BibitemShut {NoStop}%
\bibitem [{\citenamefont {Thomas}\ \emph {et~al.}(2008)\citenamefont {Thomas},
  \citenamefont {Morfill}, \citenamefont {Fortov}, \citenamefont {Ivlev},
  \citenamefont {Molotkov}, \citenamefont {Lipaev}, \citenamefont {Hagl},
  \citenamefont {Rothermel}, \citenamefont {Khrapak}, \citenamefont
  {Suetterlin}, \citenamefont {Rubin-Zuzic}, \citenamefont {Petrov},
  \citenamefont {Tokarev},\ and\ \citenamefont {Krikalev}}]{17}%
  \BibitemOpen
  \bibfield  {author} {\bibinfo {author} {\bibfnamefont {H.~M.}\ \bibnamefont
  {Thomas}}, \bibinfo {author} {\bibfnamefont {G.~E.}\ \bibnamefont {Morfill}},
  \bibinfo {author} {\bibfnamefont {V.~E.}\ \bibnamefont {Fortov}}, \bibinfo
  {author} {\bibfnamefont {A.~V.}\ \bibnamefont {Ivlev}}, \bibinfo {author}
  {\bibfnamefont {V.~I.}\ \bibnamefont {Molotkov}}, \bibinfo {author}
  {\bibfnamefont {A.~M.}\ \bibnamefont {Lipaev}}, \bibinfo {author}
  {\bibfnamefont {T.}~\bibnamefont {Hagl}}, \bibinfo {author} {\bibfnamefont
  {H.}~\bibnamefont {Rothermel}}, \bibinfo {author} {\bibfnamefont {S.~A.}\
  \bibnamefont {Khrapak}}, \bibinfo {author} {\bibfnamefont {R.~K.}\
  \bibnamefont {Suetterlin}}, \bibinfo {author} {\bibfnamefont
  {M.}~\bibnamefont {Rubin-Zuzic}}, \bibinfo {author} {\bibfnamefont {O.~F.}\
  \bibnamefont {Petrov}}, \bibinfo {author} {\bibfnamefont {V.~I.}\
  \bibnamefont {Tokarev}}, \ and\ \bibinfo {author} {\bibfnamefont {S.~K.}\
  \bibnamefont {Krikalev}},\ }\href@noop {} {\bibfield  {journal} {\bibinfo
  {journal} {New J.\ Phys.}\ }\textbf {\bibinfo {volume} {10}},\ \bibinfo
  {pages} {033036} (\bibinfo {year} {2008})}\BibitemShut {NoStop}%
\bibitem [{\citenamefont {Schwabe}\ \emph {et~al.}(2011)\citenamefont
  {Schwabe}, \citenamefont {Jiang}, \citenamefont {Zhdanov}, \citenamefont
  {Hagl}, \citenamefont {Huber}, \citenamefont {Ivlev}, \citenamefont {Lipaev},
  \citenamefont {Molotkov}, \citenamefont {Naumkin}, \citenamefont
  {S{\"{u}}utterlin}, \citenamefont {Thomas}, \citenamefont {Fortov},
  \citenamefont {Morfill}, \citenamefont {Skvortsov},\ and\ \citenamefont
  {Volkov}}]{18}%
  \BibitemOpen
  \bibfield  {author} {\bibinfo {author} {\bibfnamefont {M.}~\bibnamefont
  {Schwabe}}, \bibinfo {author} {\bibfnamefont {K.}~\bibnamefont {Jiang}},
  \bibinfo {author} {\bibfnamefont {S.}~\bibnamefont {Zhdanov}}, \bibinfo
  {author} {\bibfnamefont {T.}~\bibnamefont {Hagl}}, \bibinfo {author}
  {\bibfnamefont {P.}~\bibnamefont {Huber}}, \bibinfo {author} {\bibfnamefont
  {A.~V.}\ \bibnamefont {Ivlev}}, \bibinfo {author} {\bibfnamefont {A.~M.}\
  \bibnamefont {Lipaev}}, \bibinfo {author} {\bibfnamefont {V.~I.}\
  \bibnamefont {Molotkov}}, \bibinfo {author} {\bibfnamefont {V.~N.}\
  \bibnamefont {Naumkin}}, \bibinfo {author} {\bibfnamefont {K.~R.}\
  \bibnamefont {S{\"{u}}utterlin}}, \bibinfo {author} {\bibfnamefont {H.~M.}\
  \bibnamefont {Thomas}}, \bibinfo {author} {\bibfnamefont {V.~E.}\
  \bibnamefont {Fortov}}, \bibinfo {author} {\bibfnamefont {G.~E.}\
  \bibnamefont {Morfill}}, \bibinfo {author} {\bibfnamefont {A.}~\bibnamefont
  {Skvortsov}}, \ and\ \bibinfo {author} {\bibfnamefont {S.}~\bibnamefont
  {Volkov}},\ }\href@noop {} {\bibfield  {journal} {\bibinfo  {journal} {EPL}\
  }\textbf {\bibinfo {volume} {96}},\ \bibinfo {pages} {55001} (\bibinfo {year}
  {2011})}\BibitemShut {NoStop}%
\bibitem [{\citenamefont {Fortov}\ \emph
  {et~al.}(2004{\natexlab{b}})\citenamefont {Fortov}, \citenamefont {Petrov},
  \citenamefont {Usachev},\ and\ \citenamefont {Zobnin}}]{21}%
  \BibitemOpen
  \bibfield  {author} {\bibinfo {author} {\bibfnamefont {V.~E.}\ \bibnamefont
  {Fortov}}, \bibinfo {author} {\bibfnamefont {O.~F.}\ \bibnamefont {Petrov}},
  \bibinfo {author} {\bibfnamefont {A.~D.}\ \bibnamefont {Usachev}}, \ and\
  \bibinfo {author} {\bibfnamefont {A.~V.}\ \bibnamefont {Zobnin}},\
  }\href@noop {} {\bibfield  {journal} {\bibinfo  {journal} {Phys.\ Rev.\ E}\
  }\textbf {\bibinfo {volume} {70}},\ \bibinfo {pages} {046415} (\bibinfo
  {year} {2004}{\natexlab{b}})}\BibitemShut {NoStop}%
\bibitem [{\citenamefont {Chang}\ \emph {et~al.}(2011)\citenamefont {Chang},
  \citenamefont {Tseng},\ and\ \citenamefont {I}}]{22}%
  \BibitemOpen
  \bibfield  {author} {\bibinfo {author} {\bibfnamefont {M.-C.}\ \bibnamefont
  {Chang}}, \bibinfo {author} {\bibfnamefont {Y.-P.}\ \bibnamefont {Tseng}}, \
  and\ \bibinfo {author} {\bibfnamefont {L.}~\bibnamefont {I}},\ }\href@noop {}
  {\bibfield  {journal} {\bibinfo  {journal} {Phys.\ of Plasmas}\ }\textbf
  {\bibinfo {volume} {18}},\ \bibinfo {pages} {033704} (\bibinfo {year}
  {2011})}\BibitemShut {NoStop}%
\bibitem [{\citenamefont {Nosenko}\ and\ \citenamefont {Goree}(2004)}]{23}%
  \BibitemOpen
  \bibfield  {author} {\bibinfo {author} {\bibfnamefont {V.}~\bibnamefont
  {Nosenko}}\ and\ \bibinfo {author} {\bibfnamefont {J.}~\bibnamefont
  {Goree}},\ }\href@noop {} {\bibfield  {journal} {\bibinfo  {journal} {Phys.\
  Rev.\ Lett.}\ }\textbf {\bibinfo {volume} {93}},\ \bibinfo {pages} {155004}
  (\bibinfo {year} {2004})}\BibitemShut {NoStop}%
\bibitem [{\citenamefont {Morfill}\ \emph {et~al.}(2004)\citenamefont
  {Morfill}, \citenamefont {Rubin-Zuzic}, \citenamefont {Rothermel},
  \citenamefont {Ivlev}, \citenamefont {Klumov}, \citenamefont {Thomas},
  \citenamefont {Konopka},\ and\ \citenamefont {Steinberg}}]{24}%
  \BibitemOpen
  \bibfield  {author} {\bibinfo {author} {\bibfnamefont {G.~E.}\ \bibnamefont
  {Morfill}}, \bibinfo {author} {\bibfnamefont {M.}~\bibnamefont
  {Rubin-Zuzic}}, \bibinfo {author} {\bibfnamefont {H.}~\bibnamefont
  {Rothermel}}, \bibinfo {author} {\bibfnamefont {A.~V.}\ \bibnamefont
  {Ivlev}}, \bibinfo {author} {\bibfnamefont {B.~A.}\ \bibnamefont {Klumov}},
  \bibinfo {author} {\bibfnamefont {H.~M.}\ \bibnamefont {Thomas}}, \bibinfo
  {author} {\bibfnamefont {U.}~\bibnamefont {Konopka}}, \ and\ \bibinfo
  {author} {\bibfnamefont {V.}~\bibnamefont {Steinberg}},\ }\href@noop {}
  {\bibfield  {journal} {\bibinfo  {journal} {Phys.\ Rev.\ Lett.}\ }\textbf
  {\bibinfo {volume} {92}},\ \bibinfo {pages} {175004} (\bibinfo {year}
  {2004})}\BibitemShut {NoStop}%
\bibitem [{\citenamefont {Ivlev}\ \emph {et~al.}(2007)\citenamefont {Ivlev},
  \citenamefont {Steinberg}, \citenamefont {Kompaneets}, \citenamefont
  {H{\"{o}}fner}, \citenamefont {Sidorenko},\ and\ \citenamefont
  {Morfill}}]{Ivlev07}%
  \BibitemOpen
  \bibfield  {author} {\bibinfo {author} {\bibfnamefont {A.~V.}\ \bibnamefont
  {Ivlev}}, \bibinfo {author} {\bibfnamefont {V.}~\bibnamefont {Steinberg}},
  \bibinfo {author} {\bibfnamefont {R.}~\bibnamefont {Kompaneets}}, \bibinfo
  {author} {\bibfnamefont {H.}~\bibnamefont {H{\"{o}}fner}}, \bibinfo {author}
  {\bibfnamefont {I.}~\bibnamefont {Sidorenko}}, \ and\ \bibinfo {author}
  {\bibfnamefont {G.~E.}\ \bibnamefont {Morfill}},\ }\href@noop {} {\bibfield
  {journal} {\bibinfo  {journal} {Phys.\ Rev.\ Lett.}\ }\textbf {\bibinfo
  {volume} {98}},\ \bibinfo {pages} {145003} (\bibinfo {year}
  {2007})}\BibitemShut {NoStop}%
\bibitem [{\citenamefont {Landau}\ and\ \citenamefont {Lifshits}(1959)}]{25}%
  \BibitemOpen
  \bibfield  {author} {\bibinfo {author} {\bibfnamefont {L.~D.}\ \bibnamefont
  {Landau}}\ and\ \bibinfo {author} {\bibfnamefont {E.~M.}\ \bibnamefont
  {Lifshits}},\ }\href@noop {} {\emph {\bibinfo {title} {Fluid Mechanics}}}\
  (\bibinfo  {publisher} {Pergamon Press},\ \bibinfo {year} {1959})\BibitemShut
  {NoStop}%
\bibitem [{\citenamefont {Goree}\ \emph {et~al.}(1999)\citenamefont {Goree},
  \citenamefont {Morfill}, \citenamefont {Tsytovich},\ and\ \citenamefont
  {Vladimirov}}]{27}%
  \BibitemOpen
  \bibfield  {author} {\bibinfo {author} {\bibfnamefont {J.}~\bibnamefont
  {Goree}}, \bibinfo {author} {\bibfnamefont {G.~E.}\ \bibnamefont {Morfill}},
  \bibinfo {author} {\bibfnamefont {V.~N.}\ \bibnamefont {Tsytovich}}, \ and\
  \bibinfo {author} {\bibfnamefont {S.~V.}\ \bibnamefont {Vladimirov}},\
  }\href@noop {} {\bibfield  {journal} {\bibinfo  {journal} {Phys.\ Rev.\ E}\
  }\textbf {\bibinfo {volume} {59}},\ \bibinfo {pages} {7055} (\bibinfo {year}
  {1999})}\BibitemShut {NoStop}%
\bibitem [{\citenamefont {Nosenko}\ \emph {et~al.}(2010)\citenamefont
  {Nosenko}, \citenamefont {Ivlev},\ and\ \citenamefont {Morfill}}]{28}%
  \BibitemOpen
  \bibfield  {author} {\bibinfo {author} {\bibfnamefont {V.}~\bibnamefont
  {Nosenko}}, \bibinfo {author} {\bibfnamefont {A.~V.}\ \bibnamefont {Ivlev}},
  \ and\ \bibinfo {author} {\bibfnamefont {G.~E.}\ \bibnamefont {Morfill}},\
  }\href@noop {} {\bibfield  {journal} {\bibinfo  {journal} {Phys.\ Plasmas}\
  }\textbf {\bibinfo {volume} {17}},\ \bibinfo {pages} {123705} (\bibinfo
  {year} {2010})}\BibitemShut {NoStop}%
\bibitem [{\citenamefont {Schweigert}\ \emph {et~al.}(2002)\citenamefont
  {Schweigert}, \citenamefont {Schweigert}, \citenamefont {Nosenko},\ and\
  \citenamefont {Goree}}]{29}%
  \BibitemOpen
  \bibfield  {author} {\bibinfo {author} {\bibfnamefont {V.~A.}\ \bibnamefont
  {Schweigert}}, \bibinfo {author} {\bibfnamefont {I.~V.}\ \bibnamefont
  {Schweigert}}, \bibinfo {author} {\bibfnamefont {V.}~\bibnamefont {Nosenko}},
  \ and\ \bibinfo {author} {\bibfnamefont {J.}~\bibnamefont {Goree}},\
  }\href@noop {} {\bibfield  {journal} {\bibinfo  {journal} {Phys.\ Plasmas}\
  }\textbf {\bibinfo {volume} {9}},\ \bibinfo {pages} {4465} (\bibinfo {year}
  {2002})}\BibitemShut {NoStop}%
\bibitem [{\citenamefont {Liu}\ \emph {et~al.}(2003)\citenamefont {Liu},
  \citenamefont {Goree}, \citenamefont {Nosenko},\ and\ \citenamefont
  {Boufendi}}]{30}%
  \BibitemOpen
  \bibfield  {author} {\bibinfo {author} {\bibfnamefont {B.}~\bibnamefont
  {Liu}}, \bibinfo {author} {\bibfnamefont {J.}~\bibnamefont {Goree}}, \bibinfo
  {author} {\bibfnamefont {V.}~\bibnamefont {Nosenko}}, \ and\ \bibinfo
  {author} {\bibfnamefont {L.}~\bibnamefont {Boufendi}},\ }\href@noop {}
  {\bibfield  {journal} {\bibinfo  {journal} {Phys.\ Plasmas}\ }\textbf
  {\bibinfo {volume} {10}},\ \bibinfo {pages} {9} (\bibinfo {year}
  {2003})}\BibitemShut {NoStop}%
\bibitem [{\citenamefont {Buttensch{\"{o}}n}\ \emph {et~al.}(2011)\citenamefont
  {Buttensch{\"{o}}n}, \citenamefont {Himpel},\ and\ \citenamefont
  {Melzer}}]{31}%
  \BibitemOpen
  \bibfield  {author} {\bibinfo {author} {\bibfnamefont {B.}~\bibnamefont
  {Buttensch{\"{o}}n}}, \bibinfo {author} {\bibfnamefont {M.}~\bibnamefont
  {Himpel}}, \ and\ \bibinfo {author} {\bibfnamefont {A.}~\bibnamefont
  {Melzer}},\ }\href@noop {} {\bibfield  {journal} {\bibinfo  {journal} {New
  J.\ Phys.}\ }\textbf {\bibinfo {volume} {13}},\ \bibinfo {pages} {023042}
  (\bibinfo {year} {2011})}\BibitemShut {NoStop}%
\bibitem [{\citenamefont {Raizer}(1991)}]{RaizerBook}%
  \BibitemOpen
  \bibfield  {author} {\bibinfo {author} {\bibfnamefont {Y.~P.}\ \bibnamefont
  {Raizer}},\ }\href@noop {} {\emph {\bibinfo {title} {Gas Discharge
  Physics}}}\ (\bibinfo  {publisher} {Springer},\ \bibinfo {address} {Berlin},\
  \bibinfo {year} {1991})\BibitemShut {NoStop}%
\bibitem [{\citenamefont {Land}\ and\ \citenamefont {Goedheer}(2006)}]{Land06}%
  \BibitemOpen
  \bibfield  {author} {\bibinfo {author} {\bibfnamefont {V.}~\bibnamefont
  {Land}}\ and\ \bibinfo {author} {\bibfnamefont {W.~J.}\ \bibnamefont
  {Goedheer}},\ }\href@noop {} {\bibfield  {journal} {\bibinfo  {journal} {New
  J.\ Phys.}\ }\textbf {\bibinfo {volume} {8}},\ \bibinfo {pages} {8} (\bibinfo
  {year} {2006})}\BibitemShut {NoStop}%
\bibitem [{\citenamefont {S{\"{u}}tterlin}\ \emph {et~al.}(2009)\citenamefont
  {S{\"{u}}tterlin}, \citenamefont {Wysocki}, \citenamefont {Ivlev},
  \citenamefont {R{\"{a}}th}, \citenamefont {Thomas}, \citenamefont
  {Rubin-Zuzic}, \citenamefont {Goedheer}, \citenamefont {Fortov},
  \citenamefont {Lipaev}, \citenamefont {Molotkov}, \citenamefont {Petrov},
  \citenamefont {Morfill},\ and\ \citenamefont {L{\"{o}}wen}}]{Sutterlin09}%
  \BibitemOpen
  \bibfield  {author} {\bibinfo {author} {\bibfnamefont {K.~R.}\ \bibnamefont
  {S{\"{u}}tterlin}}, \bibinfo {author} {\bibfnamefont {A.}~\bibnamefont
  {Wysocki}}, \bibinfo {author} {\bibfnamefont {A.~V.}\ \bibnamefont {Ivlev}},
  \bibinfo {author} {\bibfnamefont {C.}~\bibnamefont {R{\"{a}}th}}, \bibinfo
  {author} {\bibfnamefont {H.~M.}\ \bibnamefont {Thomas}}, \bibinfo {author}
  {\bibfnamefont {M.}~\bibnamefont {Rubin-Zuzic}}, \bibinfo {author}
  {\bibfnamefont {W.~J.}\ \bibnamefont {Goedheer}}, \bibinfo {author}
  {\bibfnamefont {V.~E.}\ \bibnamefont {Fortov}}, \bibinfo {author}
  {\bibfnamefont {A.~M.}\ \bibnamefont {Lipaev}}, \bibinfo {author}
  {\bibfnamefont {V.~I.}\ \bibnamefont {Molotkov}}, \bibinfo {author}
  {\bibfnamefont {O.~F.}\ \bibnamefont {Petrov}}, \bibinfo {author}
  {\bibfnamefont {G.~E.}\ \bibnamefont {Morfill}}, \ and\ \bibinfo {author}
  {\bibfnamefont {H.}~\bibnamefont {L{\"{o}}wen}},\ }\href@noop {} {\bibfield
  {journal} {\bibinfo  {journal} {Phys.\ Rev.\ Lett.}\ }\textbf {\bibinfo
  {volume} {102}},\ \bibinfo {pages} {085003} (\bibinfo {year}
  {2009})}\BibitemShut {NoStop}%
\bibitem [{\citenamefont {Saigo}\ and\ \citenamefont
  {Hamaguchi}(2002)}]{Saigo02}%
  \BibitemOpen
  \bibfield  {author} {\bibinfo {author} {\bibfnamefont {T.}~\bibnamefont
  {Saigo}}\ and\ \bibinfo {author} {\bibfnamefont {S.}~\bibnamefont
  {Hamaguchi}},\ }\href@noop {} {\bibfield  {journal} {\bibinfo  {journal}
  {Phys.\ Plasmas}\ }\textbf {\bibinfo {volume} {9}},\ \bibinfo {pages} {1210}
  (\bibinfo {year} {2002})}\BibitemShut {NoStop}%
\end{thebibliography}
\end{document}